\documentclass[showpacs,twocolumn,prl,aps]{revtex4}
\usepackage{amsmath}
\usepackage{graphicx}
\usepackage{bm}

\begin{document}

\title[Short title for running header]{A continuous family of fully-frustrated Heisenberg model on the Kagome lattice}
\author{Tao Li}
\affiliation{Department of Physics, Renmin University of China,
Beijing 100872, P.R.China}
\date{\today}

\begin{abstract}
We find that the antiferromagnetic Heisenberg model on the Kagome lattice with nearest neighboring exchange coupling(NN-KAFH) belongs to a continuous family of fully-frustrated Heisenberg model on the Kagome lattice, which has no preferred classical ordering pattern. The model within this family consists of the first, second and the third neighboring exchange coupling $J_{1}$, $J_{2}$, and $J_{3}$, with $J_{2}=J_{3}$. We find that when $-J_{1}\leq J_{2}=J_{3}\leq 0.2J_{1}$, the lowest band of $J(\mathbf{q})$, namely,  the Fourier transform of the exchange coupling, is totally non-dispersive. Exact diagonalization calculation indicates that the ground state of the spin-$\frac{1}{2}$ NN-KAFH is locally stable under the perturbation of $J_{2}$ and $J_{3}$ when and only when $J_{2}=J_{3}$. Interestingly, we find that the same flat band physics is also playing an important role in the RVB description of the spin liquid state on the Kagome lattice. In particular, we show that the extensively studied $U(1)$ Dirac spin liquid state on the Kagome lattice can actually be generated from a continuous family of gauge inequivalent RVB mean field ansatz, which host very different mean field spinon dispersion. 
\end{abstract}


\maketitle

The study of the spin-$\frac{1}{2}$ Kagome antiferromagnetic Heisenberg model(KAFH) has attracted a lot of attention both theoretically and experimentally. It is generally believed that the ground state of the spin-$\frac{1}{2}$ KAFH with nearest-neighboring exchange coupling(NN-KAFH) is a quantum spin liquid state\cite{ED1,ED2,ED3,HTSE1,ED4,ED5}. However, the exact nature of such an exotic state of matter is still elusive\cite{HTSE2,TRG1}. While DMRG studies tend to imply a fully gapped $Z_{2}$ spin liquid ground state\cite{DMRG1,DMRG2,DMRG3,DMRG4}, variational studies\cite{VMC1,VMC2,VMC3,VMC4,VMC5,VMC6} and tensor network simulations\cite{TRG2} seem to prefer a gapless spin liquid ground state. At the same time, the origin of the massive number of spin singlet excitation below the spin triplet gap as found in exact diagonalization studies is still poorly understood\cite{Singlet1,Singlet2,Singlet3,Singlet4}. These problems have motivated several theoretical suggestions that the spin-$\frac{1}{2}$ NN-KAFH may sit at or be very close to a quantum critical point\cite{QCP1,QCP2,QCP3}, where two or even a massive number of phases meet. Perturbation away from the spin-$\frac{1}{2}$ NN-KAFH may thus be an illuminating way to elucidate the physics of the spin-$\frac{1}{2}$ KAFH in general. For example, DMRG studies have found evidence for a chiral spin liquid state when there is more extended exchange couplings\cite{DMRG5,DMRG6,DMRG7,DMRG8}. More recently, it is found that an exact solvable extension of the spin-$\frac{1}{2}$ NN-KAFH with an anisotropic spin exchange may hold the key to understand the intriguing relationship between the multiple phases meeting at the  spin-$\frac{1}{2}$ NN-KAFH point in the Hamiltonian parameter space\cite{Changlani1,Changlani2}.

At the semiclassical level, the spin-$\frac{1}{2}$ NN-KAFH is special in that it is fully frustrated in the sense that it has no preferred classical ordering pattern at all. For a general Heisenberg model of the form $H=\sum_{i,j}J_{i,j}\mathbf{S}_{i}\cdot\mathbf{S}_{j}$, the semiclassical ordering pattern of the system is determined by the Fourier transform of the exchange couplings, $J(\mathbf{q})$, which is in general a matrix for system with a complex lattice. More specifically, the wave vector of the semiclassical ordering is given by the momentum at which the lowest eigenvalue of $J(\mathbf{q})$ reaches its minimum. A model is fully frustrated when the lowest band of $J(\mathbf{q})$ is totally non-dispersive, since there is no preferred ordering wave vector any more. This is impossible on a simple lattice, when $J(\mathbf{q})$ is simply a number, since the exact flatness of $J(\mathbf{q})$ can only be achieved when $J_{i,j}=0$ for all $i\neq j$. However, for system with a complex lattice, one or more bands of $J(\mathbf{q})$ can be totally non-dispersive even if $J_{i,j}$ is nontrivial. For example, the lowest band of $J(\mathbf{q})$ is exactly flat in the case of the NN-KAFH.

\begin{figure}[h!]
\includegraphics[width=6cm,angle=0]{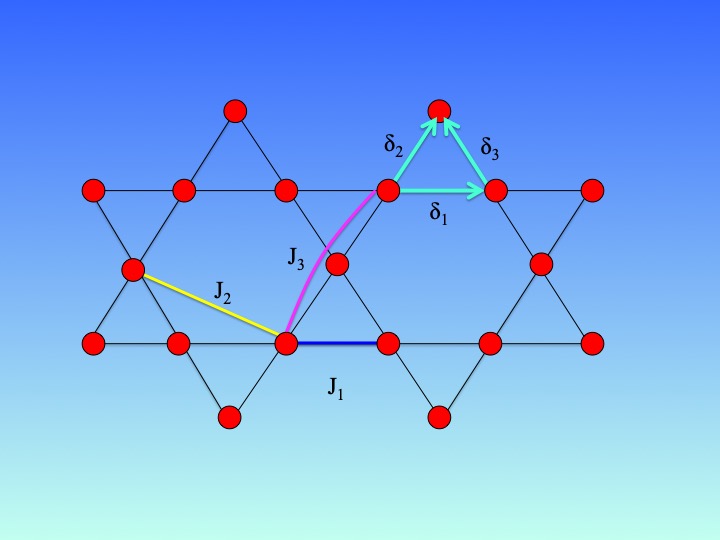}
\caption{The Kagome lattice and the exchange couplings between the first, second and the third-neighboring sites, denoted here as $J_{1}$, $J_{2}$ and $J_{3}$ respectively. $\bm{\delta}_{1}$, $\bm{\delta}_{2}$ and $\bm{\delta}_{3}$ are the three vectors connecting the first-neighboring sites of the Kagome lattice.} 
\label{fig1}
\end{figure}

In general, such fully frustrated model can be realized only when the exchange couplings in the Hamiltonian take value at isolated point in the Hamiltonian parameter space. Perturbation away from such special point will immediately lift the semiclassical degeneracy and choose for the system a particular classical ordering pattern. This is of course not what we want if our purpose is to discover more exotic quantum phases. Here we show that the NN-KAFH does not correspond to such an isolated point in the space of Hamiltonian parameters, but is within a continuous family of fully frustrated models. More specifically, we show that the flat band in the spectrum of $J(\mathbf{q})$ is robust against the introduction of the second and the third neighbor exchange coupling, namely, $J_{2}$ and $J_{3}$ as illustrated in Fig.1, provided that $J_{2}=J_{3}$. We find that for $-J_{1}<J_{2}=J_{3}<0.2J_{1}$, the flat band of $J(\mathbf{q})$ is always the lowest band and the model is thus always fully frustrated. The study of such a continuous family of fully frustrated model may shed important light on the physics of the spin-$\frac{1}{2}$ NN-KAFH.

\begin{figure}[h!]
\includegraphics[width=6cm,angle=0]{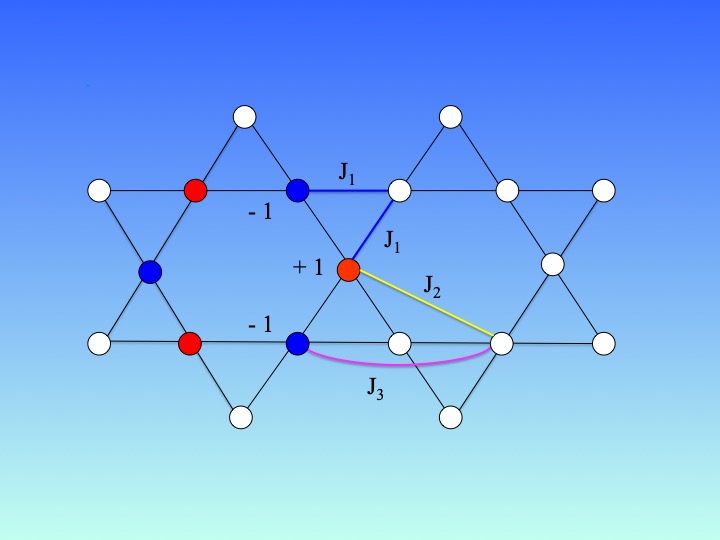}
\caption{Demonstration of the destructive interference between the hopping amplitudes out of a localized Wannier orbital defined on an elementary hexagon of the Kagome lattice. The wave function amplitudes on the red, blue and white site are $+1$, $-1$ and 0 respectively. One can easily check that the hopping amplitudes from the red and the blue sites to any given white site add to zero when $J_{2}=J_{3}$.} \label{fig2}
\end{figure}

The $J(\mathbf{q})$ of the NN-KAFH has the same form as the Hamiltonian matrix of a free electron with hoping integral $J_{1}$ between nearest neighboring sites on the Kagome lattice and is given by
\begin{equation}
J(\mathbf{q})=2J_{1}\left(\begin{array}{ccc} 0 & \cos q_{1}&  \cos q_{2}\\ \cos q_{1}&0 &\cos q_{3} \\ \cos q_{2}&\cos q_{3}&0 \end{array}\right),
\end{equation}
in which $q_{i}=\mathbf{q}\cdot \bm{\delta}_{i}$. $\bm{\delta}_{1}$,$\bm{\delta}_{2}$ and $\bm{\delta}_{3}$ are the three vectors connecting nearest neighboring sites of the Kagome lattice(see Fig. 1 for an illustration).  Using the identity $\bm{\delta}_{3}=\bm{\delta}_{2}-\bm{\delta}_{1}$, one can easily prove the existence of a flat band in the spectrum of $J(\mathbf{q})$ with an eigenvalue of $-2J_{1}$. The origin of this flat band can be understood more intuitively in real space. More specifically, it can be attributed to the destructive interference between the hopping amplitudes out of a localized Wannier orbital defined on an elementary hexagon of the Kagome lattice\cite{Flat}, as is illustrated in Fig.2. Interestingly, one find that the destructive interference remians effective even if we introduce the second and the third neighbor exchange coupling, provided that $J_{2}=J_{3}$. We thus expect the flat band to be robust against the introduction of $J_{2}$ and $J_{3}$, provided that $J_{2}=J_{3}$. Indeed, one find that $J(\mathbf{q})$ always has $-2(J_{1}-J_{2})$ as one of its three eigenvalues when $J_{2}=J_{3}$.

The KAFH is fully frustrated when the above flat band becomes the lowest band of $J(\mathbf{q})$. In Fig.3, we plot the band structure of $J(\mathbf{q})$ for several values of $J_{2}(=J_{3})$. From the plot we see that the flat band is the lowest band of $J(\mathbf{q})$ when $-J_{1}<J_{2}=J_{3}<0.2J_{1}$. The NN-KAFH is thus within a continuous family of fully frustrated models, rather than an isolated point in the space of Hamiltonian parameters. The existence of such a continuous family of fully frustrated Heisenberg model on the Kagome lattice offers a much broader playground for the search of quantum spin liquid state on the Kagome lattice. In particular, it is interesting to know to what extent the intriguing physics of the spin-$\frac{1}{2}$ NN-KAFH persist within the space of this continuous family of fully frustrated models\cite{davidstar}.

\begin{figure}[h!]
\includegraphics[width=4.2cm,height=3.5cm]{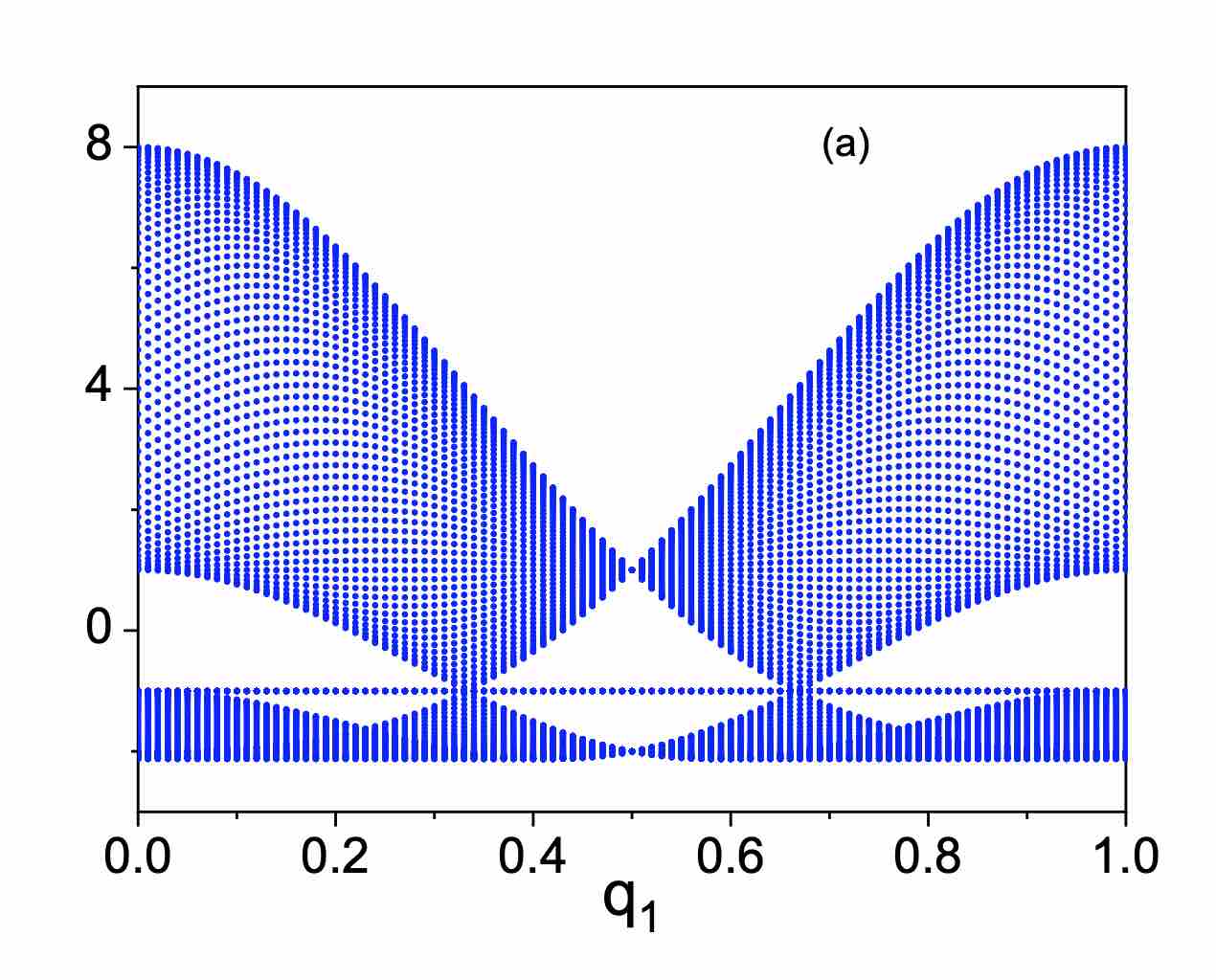}
\includegraphics[width=4.2cm,height=3.5cm]{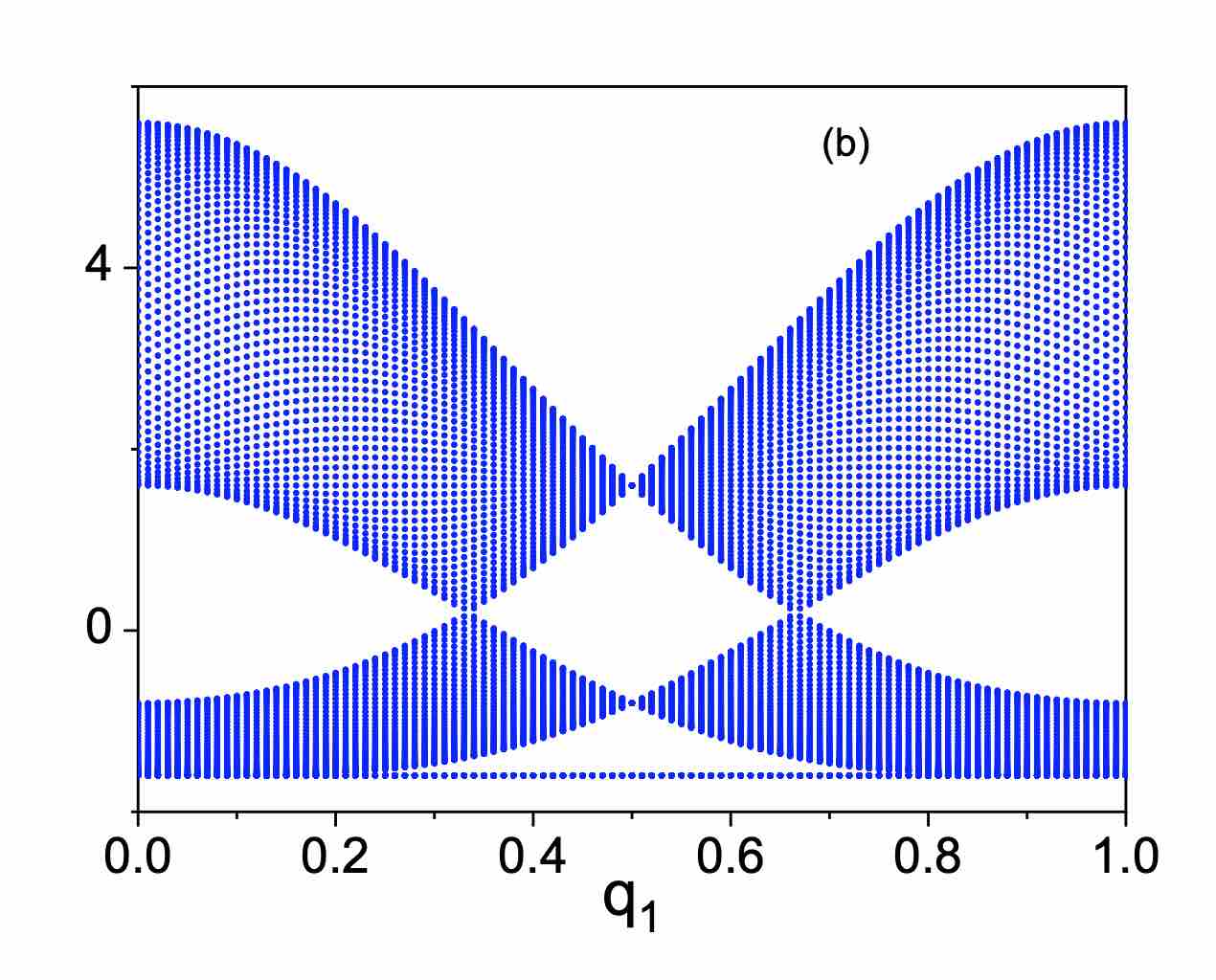}
\includegraphics[width=4.2cm,height=3.5cm]{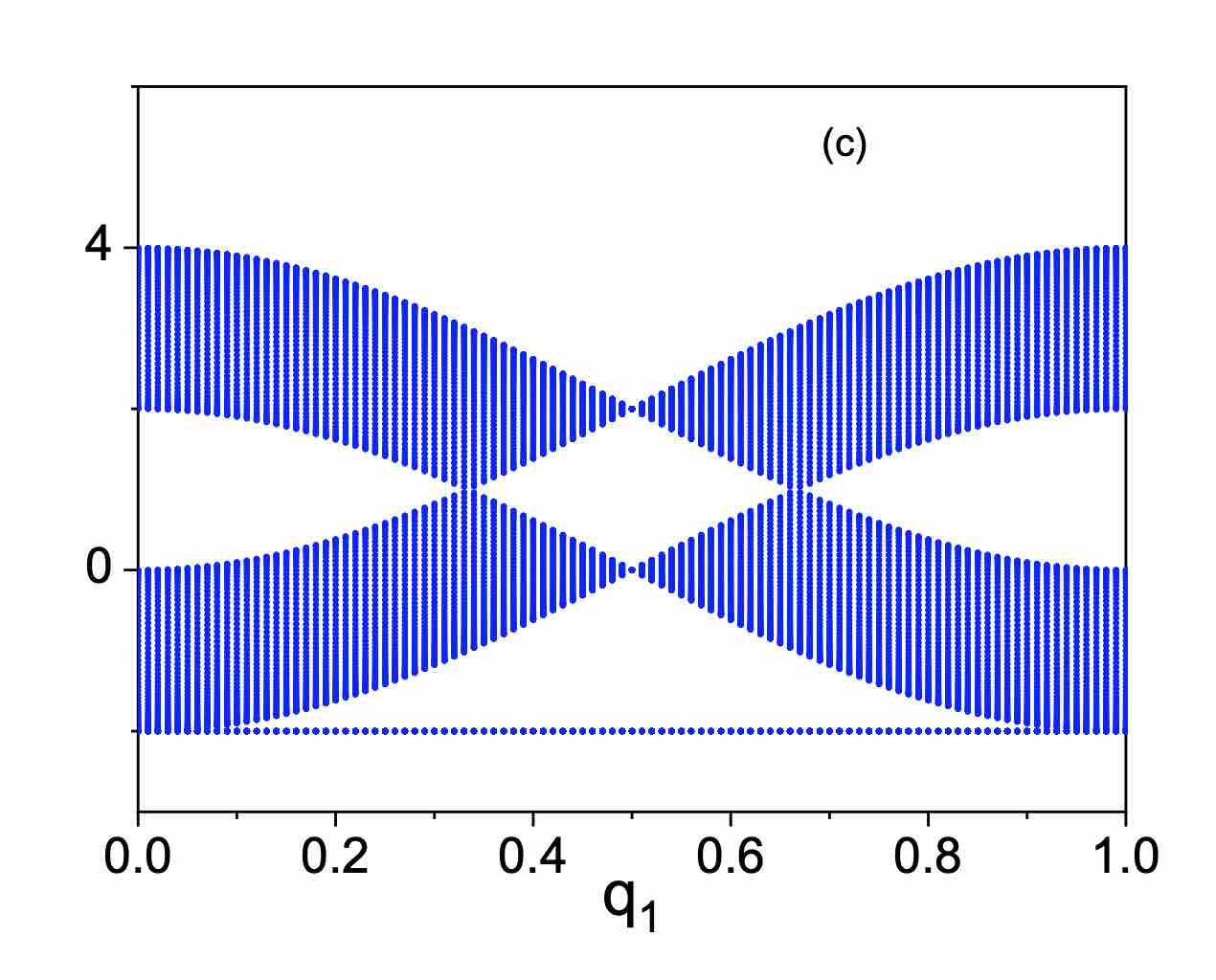}
\includegraphics[width=4.2cm,height=3.5cm]{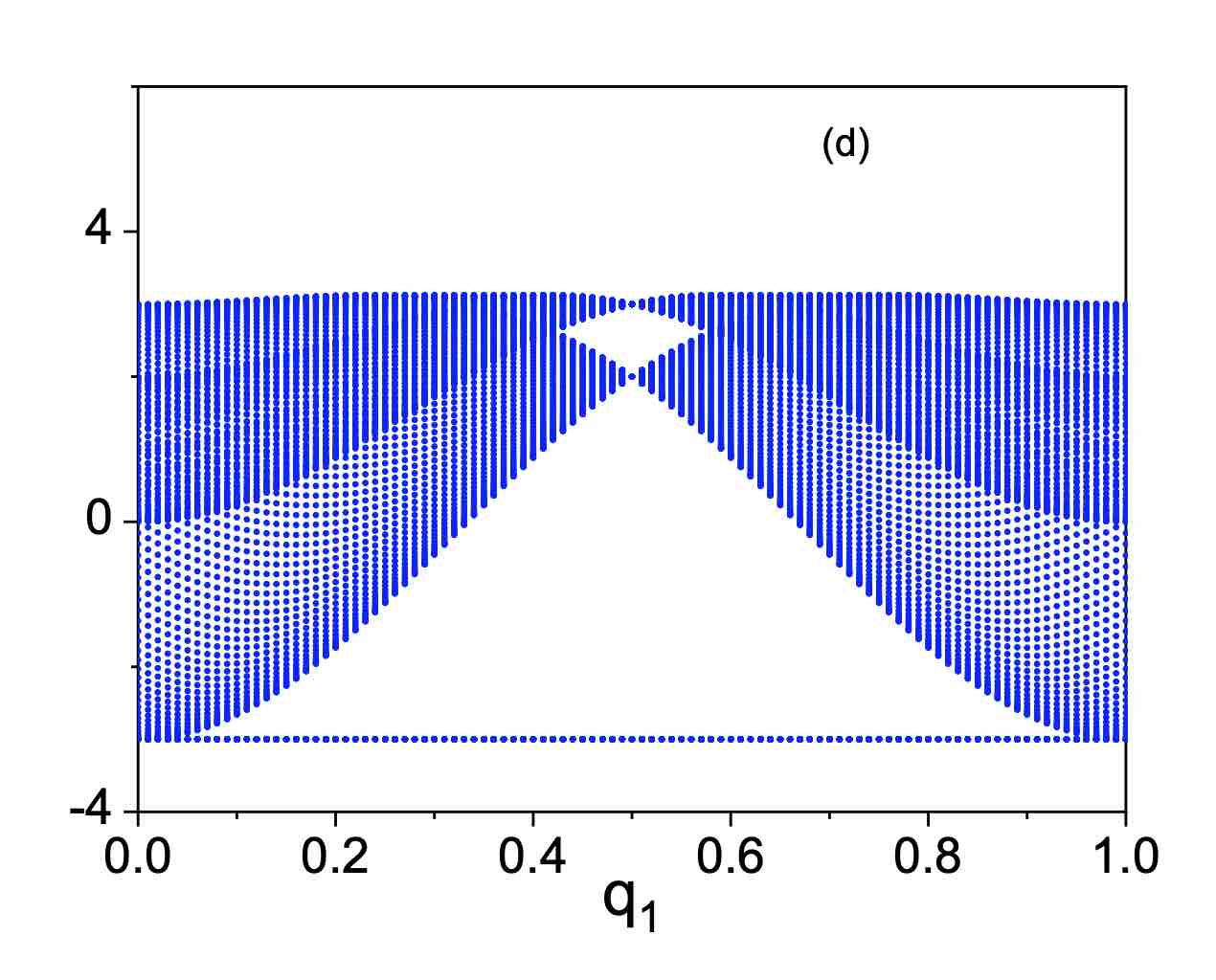}
\includegraphics[width=4.2cm,height=3.5cm]{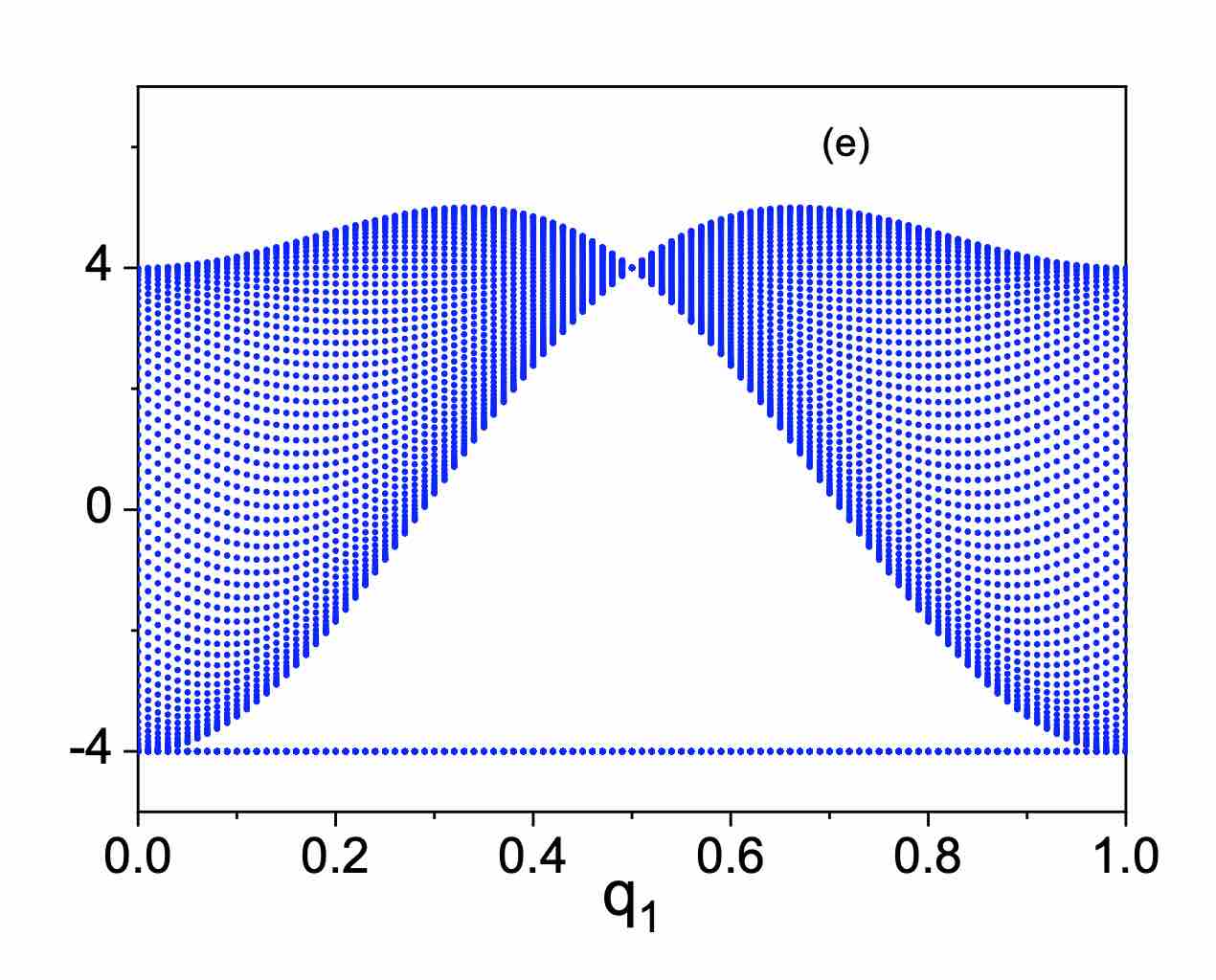}
\includegraphics[width=4.2cm,height=3.5cm]{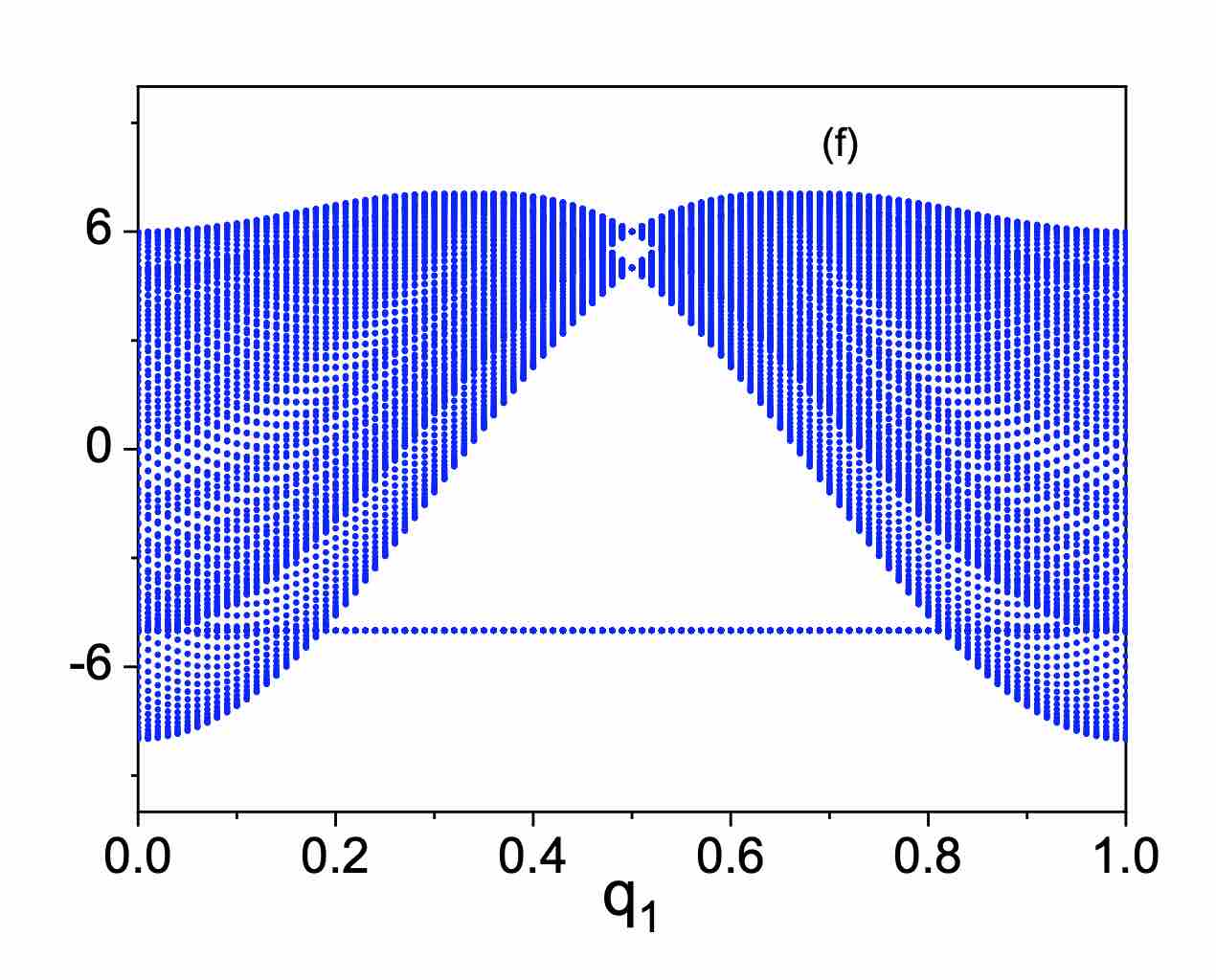}
\caption{Evolution of the band structure of $J(\mathbf{q})$ with $J_{2}(=J_{3})$. $J_{1}$ is set as the unit of energy in the plots. (a)$J_{2}=0.5$, (b)$J_{2}=0.2$, (c)$J_{2}=0$, (d)$J_{2}=-0.5$, (e)$J_{2}=-1$ and (f)$J_{2}=-1.5$. The collection of eigenvalues of $J(\mathbf{q})$ at different $q_{2}$ form three intertwining bands as a function of $q_{1}$. The flat band at $-2(J_{1}-J_{2})$ manifests itself as a straight line in the plot. It lies at the bottom of the spectrum when $-1\leq J_{2}=J_{3}\leq 0.2$. $q_{1}$ is plotted in unit of $2\pi$.} \label{fig3}
\end{figure}

\begin{figure}[h!]
\includegraphics[width=8cm,angle=0]{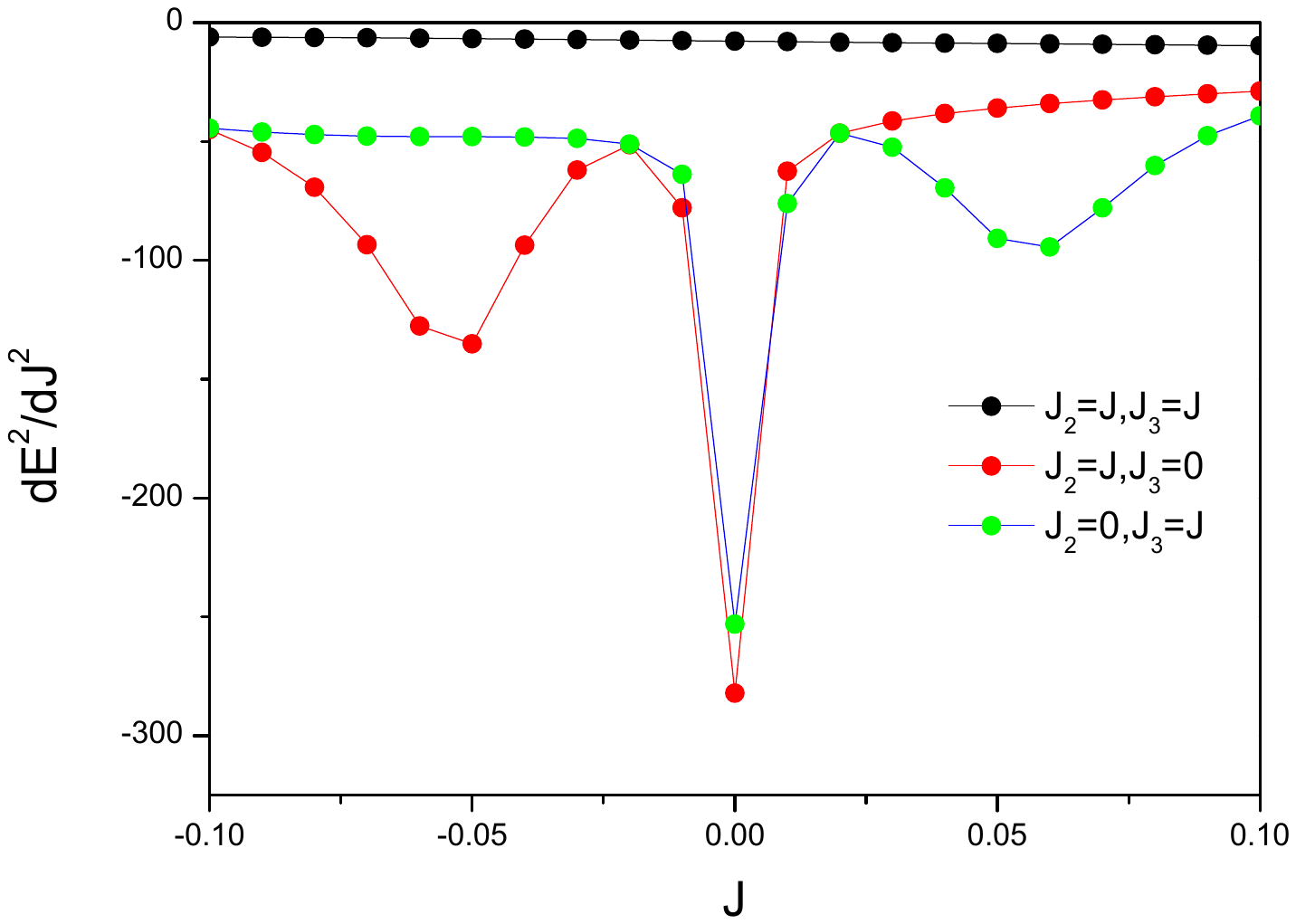}
\caption{The second derivative of the ground state energy as a function of the perturbation away from the spin-$\frac{1}{2}$ NN-KAFH point along the line $J_{2}=\alpha J_{3}$. Here we have presented the results for three typical cases, namely, when $J_{2}=J_{3}=J$, $J_{2}=J, J_{3}=0$ and $J_{2}=0, J_{3}=J$. $J_{1}$ is set as the unit of energy in the plot. We find that the huge peak in $d^{2}E/dJ^{2}$ at $J=0$ persists whenever $\alpha \neq 1$.} \label{fig4}
\end{figure}

For this reason, we have performed exact diagonalization calculation of the $J_{1}-J_{2}-J_{3}$ model on a finite cluster of 36 sites. To diagnose the stability of the ground state of the spin-$\frac{1}{2}$ NN-KAFH under the perturbation of $J_{2}$ and $J_{3}$, we have calculated the second derivative of the ground state energy with respect to the Hamiltonian parameter along the line $J_{2}=\alpha J_{3}$. We find that the ground state energy of the system behaves rather smoothly when $\alpha=1$. This is in stark contrast with the situation when $\alpha\neq 1$, for which a huge peak in $d^{2}E/dJ^{2}$ is observed at the spin-$\frac{1}{2}$ NN-KAFH point. We note that previous study find that the second derivative of the ground state energy with respect to the six-site ring exchange coupling $J_{r}$ also exhibits a sharp peak at the spin-$\frac{1}{2}$ NN-KAFH point\cite{QCP3}, which implies that the spin-$\frac{1}{2}$ NN-KAFH may sit at or be very close to a quantum critical point. The results of the present work indicate that the perturbation corresponding to $J_{2}$ and $J_{3}$ are also strongly relevant at the spin-$\frac{1}{2}$ NN-KAFH point. However, a special combination of the two along the line $J_{2}=J_{3}$ is irrelevant. This give us the hope that the spin liquid ground state of the spin-$\frac{1}{2}$ NN-KAFH may form a phase of finite range of stability in the space of Hamiltonian parameters, rather than being an isolated critical point. It is thus interesting to study the $J_{1}-J_{2}-J_{3}$ model along the line $J_{2}=J_{3}$ further with other numerical approaches.

\begin{figure}[h!]
\includegraphics[width=6cm,angle=0]{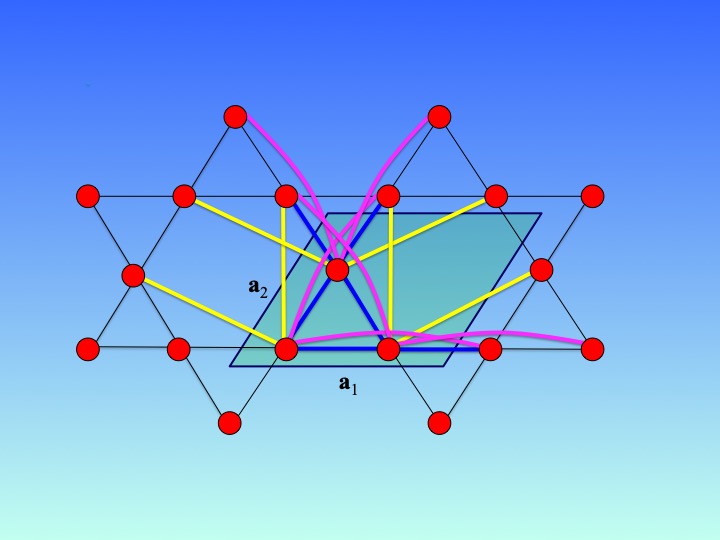}
\caption{Illustration of the RVB mean field ansatz of the $U(1)$ Dirac spin liquid state.  The green 
parallelogram denotes the unit cell of the Kagome lattice, with $\mathrm{\mathbf{a}}_{1}$ and $\mathrm{\mathbf{a}}_{2}$ as its two basis vectors.
 The spinon unit cell is doubled in the $\mathrm{\mathbf{a}}_{1}$ direction. The blue, yellow and pink lines denote the first, second and the third neighboring bonds of the Kagome lattice. $\chi_{i,j}$ is translational invariant along the $\mathrm{\mathbf{a}}_{2}$ direction, but will change sign when translated in the $\mathrm{\mathbf{a}}_{1}$ direction by one lattice constant, if the cell indices in the $\mathrm{\mathbf{a}}_{2}$ direction of site $i$ and $j$ differ by an odd number. Such a sign change is dictated by the phase factor $s_{i,j}$ in $\chi_{i,j}$. $s_{i,j}$ equals to one on the blue, yellow and pink bonds shown here.} \label{fig5}
\end{figure}

Interestingly, we find that the flat band physics discussed above is important not only for the determination of the classical ordering pattern in the KAFH, it also plays an important role in the RVB description of the spin liquid physics of the spin-$\frac{1}{2}$ KAFH. Here we will concentrate on the extensively studied $U(1)$ Dirac spin liquid on the Kagome lattice\cite{VMC1,VMC2,VMC3,VMC4}. This state is constructed from Gutzwiller projection of the ground state of the following mean field Hamiltonian(also called an RVB mean field ansatz in the literature)  
\begin{equation}
H_{MF}=\sum_{i,j,\sigma}\chi_{i,j}f_{i,\sigma}^{\dagger}f_{j,\sigma},\nonumber
\end{equation}
in which
\begin{eqnarray}
\chi_{i,j}=\left\{\begin{aligned}
                -\chi_{1}\  s_{i,j}& & \mathrm{first \ neighbor}\\  
                -\chi_{2}\  s_{i,j} & & \mathrm{second \ neighbor}\\
                -\chi_{3}\ s_{i,j} & &\mathrm{third \ neighbor}
\end{aligned}  
\right.
\end{eqnarray}
Here $s_{i,j}=\pm1$ is the phase factor introduced to ensure that a $\pi$-flux is enclosed in each unit cell of the Kagome lattice(see Fig.5). More specifically, $\chi_{i,j}$ is translational invariant in the $\mathrm{\mathbf{a}}_{2}$ direction, but will change sign when translated by one unit in the $\mathrm{\mathbf{a}}_{1}$ direction, if the unit cell indices of site $i$ and $j$ in the $\mathrm{\mathbf{a}}_{2}$ direction differ by an odd number. 

\begin{figure}
\includegraphics[width=6cm,angle=0]{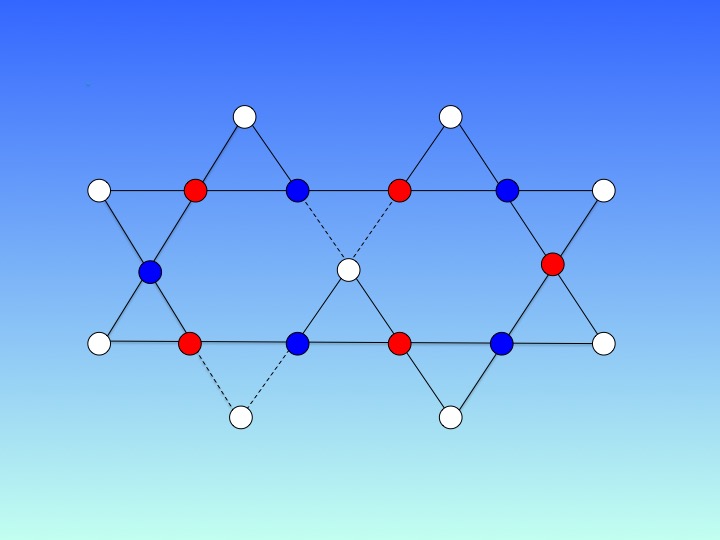}
\caption{Illustration of the localized Wannier orbital of the flat band of $H_{MF}$. The wave function amplitude on the red, blue and white sites are +1, -1 and 0 respectively. The hopping integral on the dashed bonds has an additional minus sign as a result of the phase factor $s_{i,j}$ in Eq.(2)(we have not plotted out the second and third neighbor hopping explicitly for clarity). The hopping amplitudes from the red and the blue sites to any given white site add to zero when $\chi_{2}=\chi_{3}$.} \label{fig6}
\end{figure}

Just as in the case of $J(\mathbf{q})$, a non-dispersive band at $2(\chi_{1}-\chi_{2})$ emerges in the spectrum of $H_{MF}$ when $\chi_{2}=\chi_{3}$. The origin of such a flat band can also be attributed to the destructive interference between the hopping amplitudes out of a localized Wannier orbital, which now involves two unit cell of the Kagome lattice as a result of the $\pi$-flux structure encoded in the phase factor $s_{i,j}$. An illustration of such a localized Wannier orbital is presented in Fig.6. 

More interestingly, one find that the ground state of $H_{MF}$(and thus the Gutzwiller projected RVB state) is totally unchanged when we tune the value of $\chi_{2}(=\chi_{3})$ in the range $\alpha=\frac{\chi_{2}}{\chi_{1}}\ \in[-0.6,0.27]$. This striking result can be understood by rewriting $H_{MF}$ as $H_{MF}=H_{1}+H_{2}+H_{3}$, in which $H_{1}$, $H_{2}$ and $H_{3}$ denote the part of $H_{MF}$ that is proportional to $\chi_{1}$, $\chi_{2}$ and $\chi_{3}$. It is then straightforward to check that 
$[H_{1},H_{2}+H_{3}]=0$ when $\chi_{2}=\chi_{3}$. The inclusion of $H_{2}+H_{3}$ with $\chi_{2}=\chi_{3}$ will thus only modify the eigenvalues of $H_{MF}$, but not its eigenvectors. In Fig.7, we plot the spectrum of $H_{MF}$ for different values of $\alpha$. One find that as we increase the value of $\alpha$, the flat band moves downward continuously. For $\alpha\in[-0.6,0.27]$, the flat band lies aways above the Dirac point and the occupied states is independent of the value of $\alpha$. Such a singular(non-injective) behavior in the mapping between the RVB mean field ansatz and the Gutzwiller projected RVB state will greatly complicate the optimization of the RVB parameters around the $U(1)$ Dirac spin liquid state. More specifically, the variational energy is expected to exhibit a long and narrow valley with a very flat bottom around the non-injective line $\alpha=1$ in the space of variational parameters. In a related work\cite{Tao}, we show that the best RVB state of the spin-$\frac{1}{2}$ NN-KAFH is described by a $Z_{2}$ gapped mean field ansatz in the close vicinity of the above non-injective line\cite{VMC1,VMC2,VMC3,VMC4}.

\begin{figure}[h!]
\includegraphics[width=4.2cm,height=3.5cm]{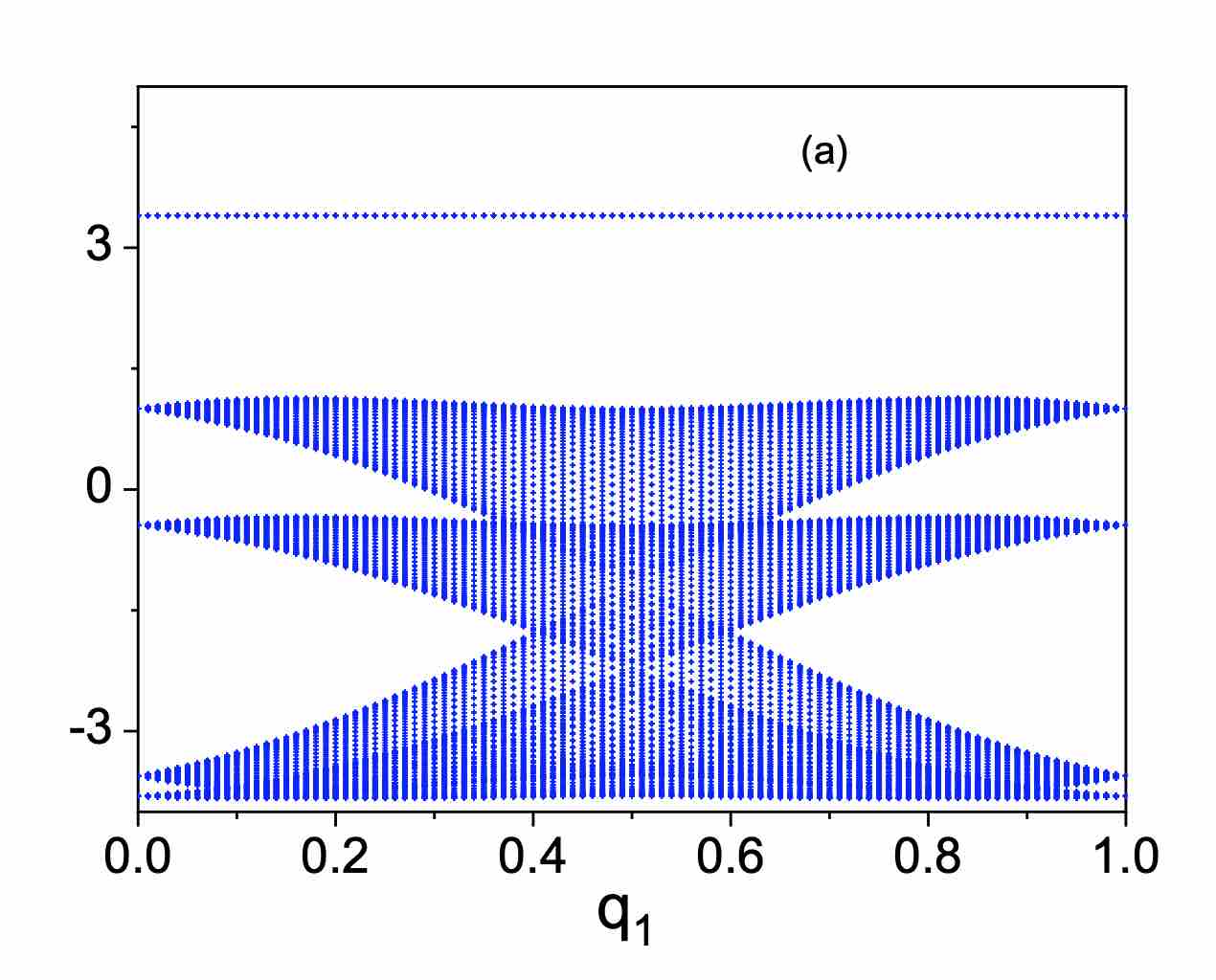}
\includegraphics[width=4.2cm,height=3.5cm]{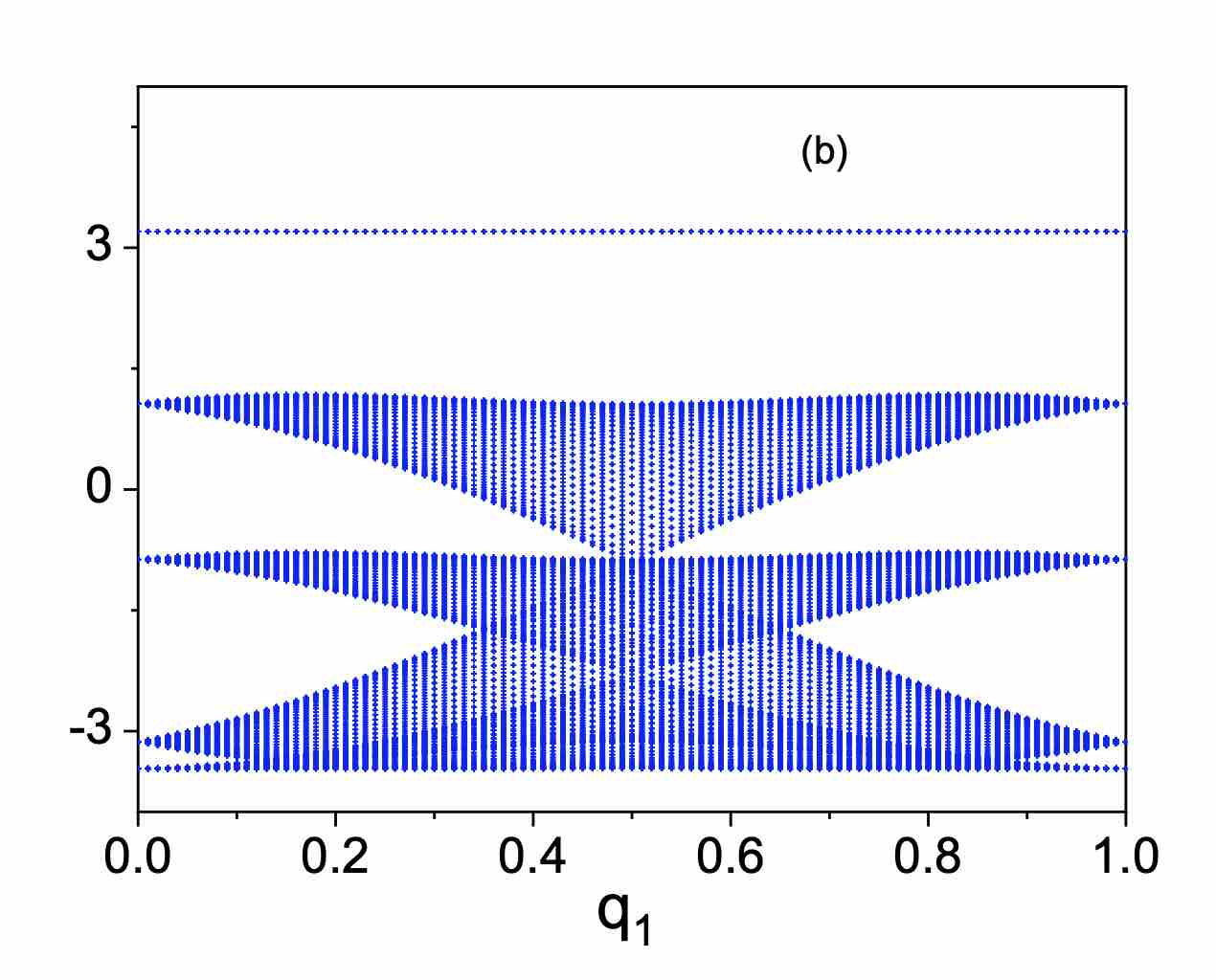}
\includegraphics[width=4.2cm,height=3.5cm]{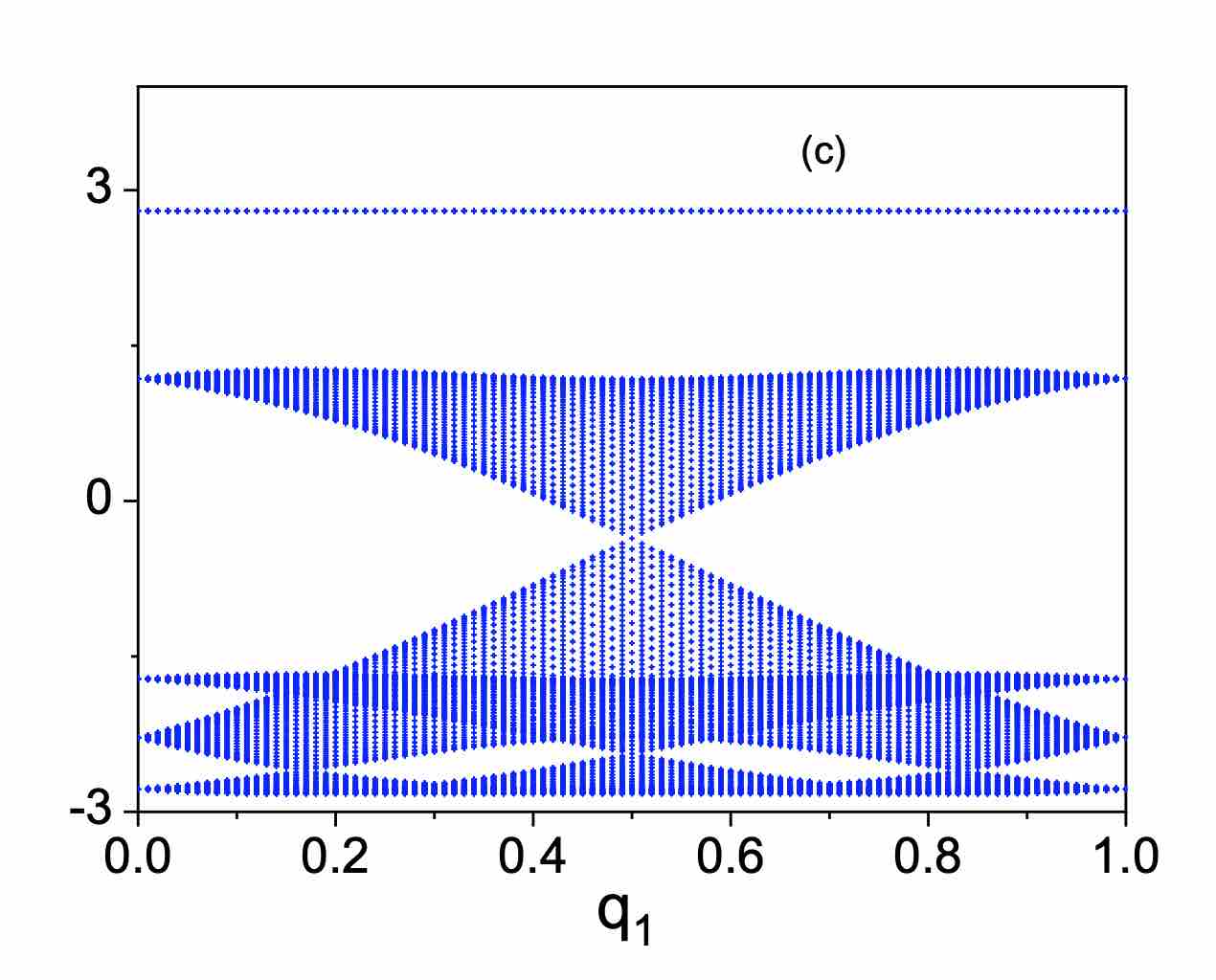}
\includegraphics[width=4.2cm,height=3.5cm]{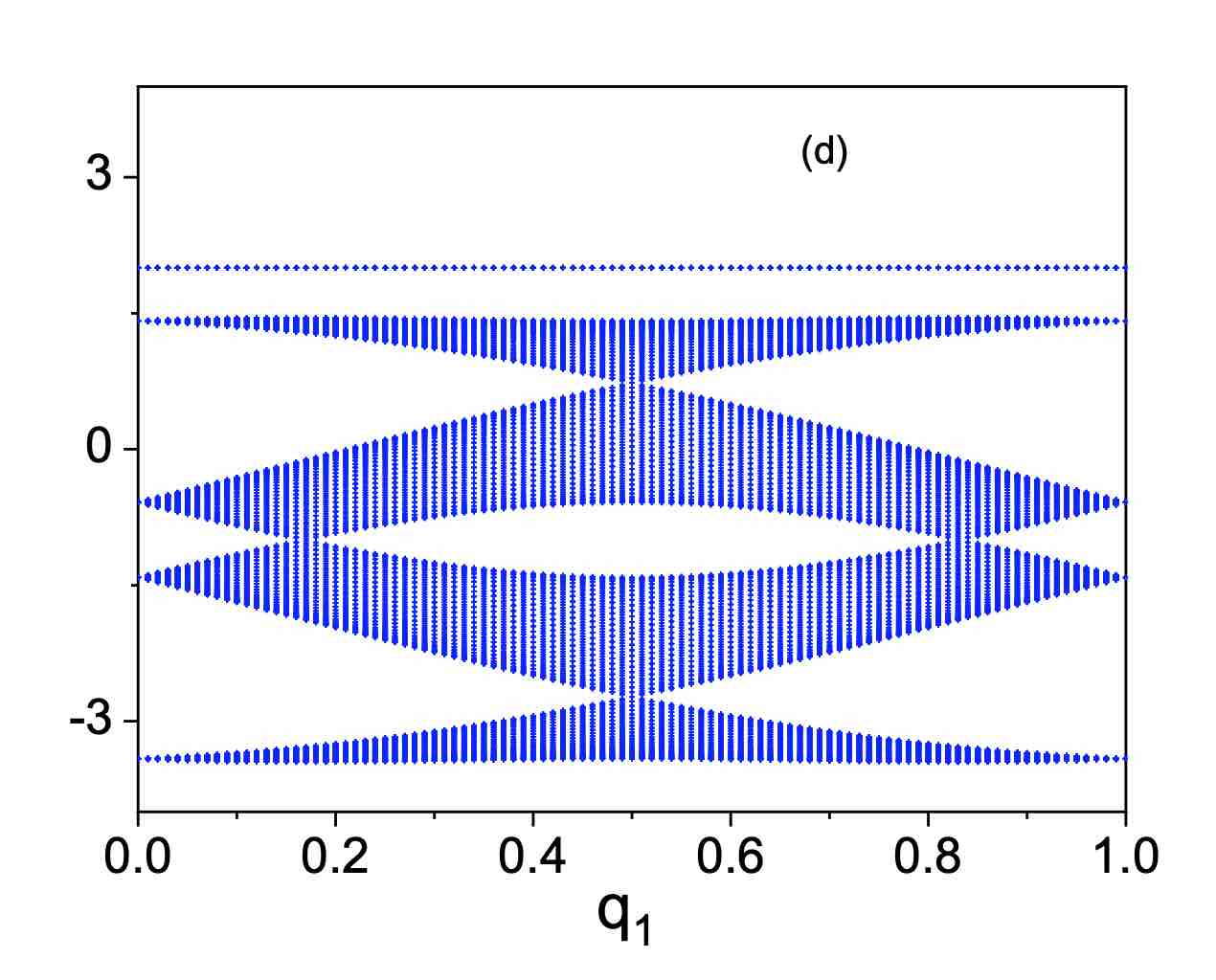}
\includegraphics[width=4.2cm,height=3.5cm]{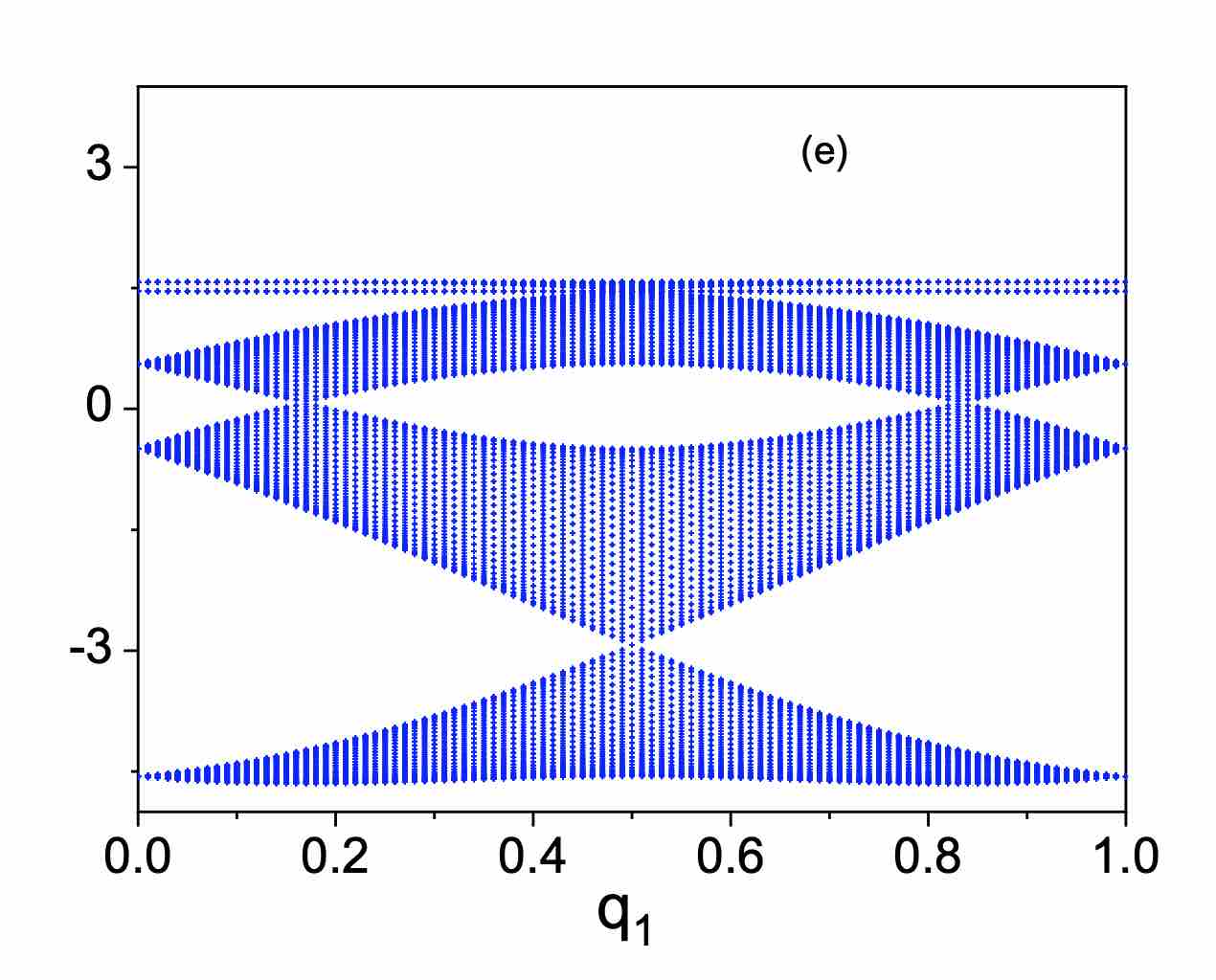}
\includegraphics[width=4.2cm,height=3.5cm]{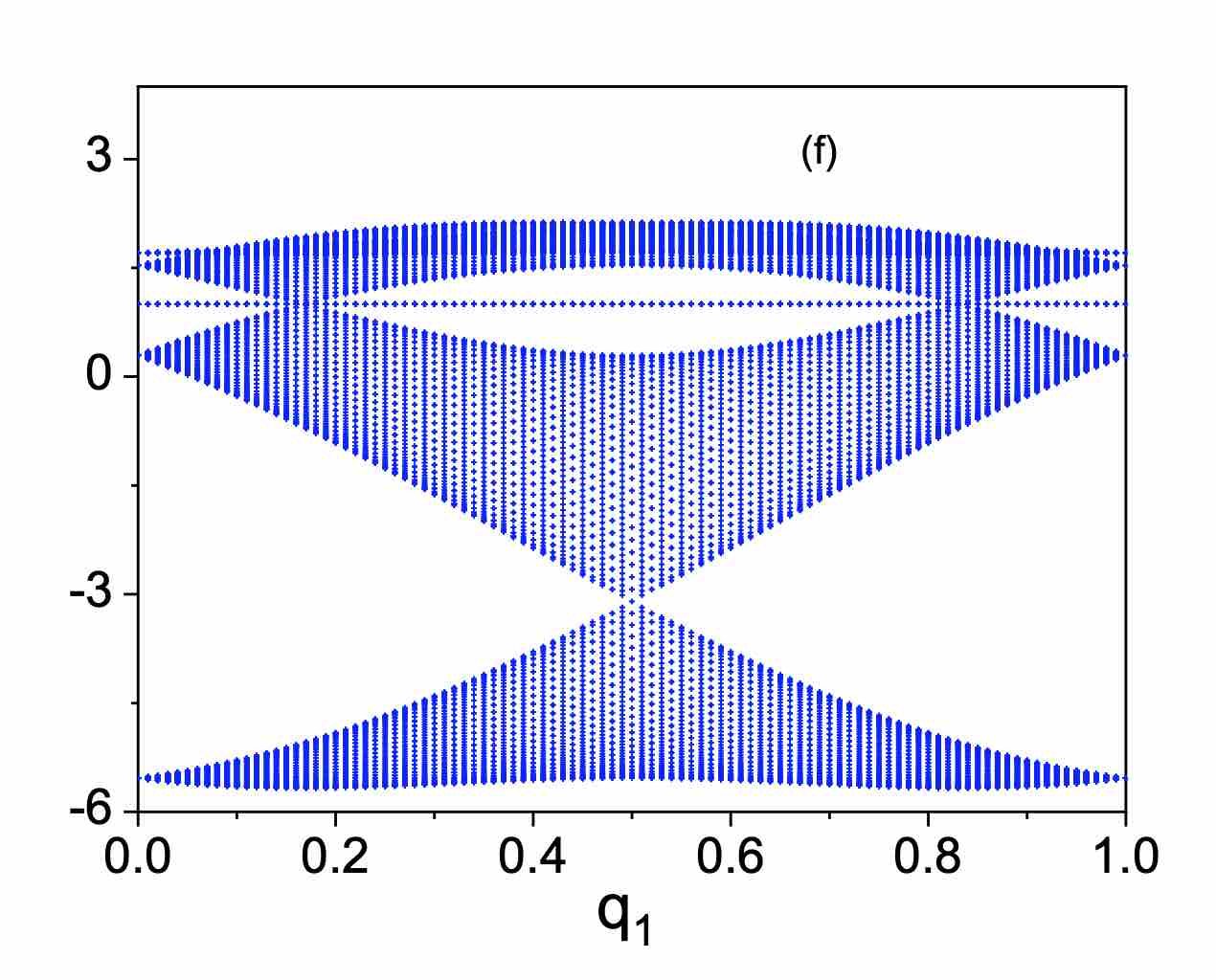}
\caption{Evolution of the spectrum of $H_{MF}$ with $\alpha=\chi_{2}(=\chi_{3})$. $\chi_{1}$ is set as the unit of energy in the plot. (a)$\alpha=-0.7$, (b)$\alpha=-0.6$, (c)$\alpha=-0.4$, (d)$\alpha=0$, (e)$\alpha=0.27$, (f)$\alpha=0.5$. The collection of eigenvalues of $H_{MF}$ at different $q_{2}$ form six intertwining bands as a function of $q_{1}$. The flat band at $2(1-\alpha)$(doubly degenerate) manifests itself as a straight line in the plot.} \label{fig6}
\end{figure}

We note that the mean field spinon dispersion predicted by $H_{MF}$ is strongly $\alpha$-dependent in the range $\alpha\in[-0.6,0.27]$. This is counter-intuitive since we usually expect that the excitation spectrum of a quantum system should be determined by its ground state structure. The lack of such a correlation in the current case indicates that the RVB mean field theory has serious problems in the description of spin liquid physics on the Kagome lattice. In a related work\cite{Tao}, we show that the ambiguity related to the RVB mean field theory can be resolved by enforcing the no double occupancy constraint on the spinon operator by Gutzwiller projection. In particular, we show that the spin fluctuation spectrum calculated from a Gutzwiller projected RPA theory on the $U(1)$ Dirac spin liquid state actually does not depends on the value of $\alpha$.

In summary, we find that there is a continuous family of fully frustrated Heisenberg models on the Kagome lattice, within which the extensively studied spin-$\frac{1}{2}$ NN-KAFH is but an ordinary point. This discovery greatly enlarge the playground for the search of exotic physics in the spin-$\frac{1}{2}$ KAFH. We find that the flat band physics that is responsible for the fully frustrated nature of the spin model is also playing an important role in the RVB description of the spin liquid physics on the Kagome lattice. In particular, we find that extensively studied $U(1)$ Dirac spin liquid can be generated from a continuous family of gauge inequivalent RVB parameters, which indicates that the RVB mean field theory has serious problem in the description of the spin liquid physics on the Kagome lattice.

We acknowledge the support from the National Natural Science Foundation of China(Grant No. 11674391), the Research Funds of Renmin University of China(Grant No.15XNLQ03), and the National Program on Key Research Project(Grant No.2016YFA0300504).


\begin{thebibliography}{99}
\bibitem{davidstar}We note that up to the third neighboring exchange coupling, there is an additional point in the parameter space at which the Heisenberg model becomes fully frustrated. This point is given by $J_{1}=J_{2}=J_{3d}$, with $J_{3d}$ the exchange coupling across the diagonal of the elementary hexagon of the Kagome lattice. The localized Wannier orbital corresponding to the flat band of this model is defined on a elementary David star of the Kagome lattice and consists of 12 sites. The wave function amplitudes on the 12 sites are given by $1$, $e^{i\frac{2\pi}{3}}$ and $e^{i\frac{4\pi}{3}}$. However, this fully frustrated point is an isolated point in the parameter space.
\end{thebibliography}
\end{document}